\documentclass[journal ]{new-aiaa}
\usepackage[utf8]{inputenc}
\usepackage{textcomp}

\usepackage{graphicx}
\usepackage{amsmath}
\usepackage[version=4]{mhchem}
\usepackage{siunitx}
\usepackage{longtable,tabularx}
\title{Non-Equilibrium Phase Changes in Aircraft Exhaust:\\ A Computational Study on Early Contrail Formation}

\author{Katharina Tegethoff\footnote{Postdoctoral Research Associate, Whittle Laboratory, Department of Engineering. Email: kt568@cam.ac.uk} and Andrew P.~S.~Wheeler\footnote{Professor of Aerothermal and Fluids Engineering, Whittle Laboratory, Department of Engineering. Email: aw329@cam.ac.uk}}
\affil{University of Cambridge, Cambridge, CB3 0DY}

\begin{document}

\maketitle

\begin{abstract}
A numerical framework is developed to model contrail formation in the near-field exhaust of aircraft engines, resolving non-equilibrium phase transitions in compressible, multi-component, non-ideal fluid flows. The approach combines well-established methods from steam turbine modeling for liquid-phase transitions with cloud microphysics models for ice formation. It resolves homogeneous and heterogeneous nucleation, interphase momentum exchange, and polydispersed size distributions of droplets, ice crystals, and soot particles. These models are implemented in a parallelized finite-volume solver and applied to a high-bypass turbofan exhaust configuration with simplified geometry. Results indicate that non-equilibrium effects strongly influence condensation and freezing dynamics, while nozzle geometry and water vapor content modulate local supersaturation and phase transition pathways. The findings underscore the limitations of equilibrium-based models and highlight the value of physics-based, scalable tools for analyzing contrail formation across fuels and propulsion systems.
\end{abstract}

\section*{Nomenclature}
{\renewcommand\arraystretch{1.0}
\noindent\begin{longtable*}{@{}l @{\quad=\quad} l@{}}
$A_{C}$ & Cunningham correction factor \\
$a_{w}$ & water activity \\
$C_{D}$ & dimensionless drag coefficient \\
$c_{p}$ & specific isobaric heat capacity [\si{\joule\per\kilogram\per\kelvin}]\\
$D$ & diffusion coefficient [\si{\meter\squared\per\second}] \\
$D_{core},D_{fan}$ & diameter of core and fan [\si{\meter}] \\
$E,e$ & total and specific inner energy [\si{\joule\per\kilogram}]\\
$\mathbf{F}^{c},\mathbf{F}^{v}$ & convective and viscous flux vector \\
$f(r)$ & number density function of radii [\si{\per\meter\tothe{4}}]\\
$f(\theta)$ & geometric factor related to surface contact angle $\theta$ \\
$G(r)$ & growth rate [\si{\meter\per\second}]\\
$g$ & specific Gibbs energy [\si{\joule\per\kilogram}]\\
$H, h, h_{fg}$ & total and specific enthalpy, specific latent heat [\si{\joule\per\kilogram}]\\
$J$ & nucleation rate [\si{\per\cubic\meter\per\second}]\\
$Kn$ & Knudsen number \\
$k_{b}$ & Boltzmann constant [\si{\meter\squared\kilogram\per\second\squared\per\kelvin}]\\
$L_{x},L_{r}$ & spatial dimensions of computational domain [\si{\meter}]\\
$\bar{l}$ & mean free path [\si{\meter}]\\
$M$ & molar mass [\si{\kilogram\per\mole}]\\
$Ma$ & Mach number \\
$m$ & mass [\si{\kilogram}]\\
$N_{A}$ & Avogadro constant [\si{\per\mole}]\\
$n$ & number of droplets/particles/crystals \\
$n_{s}$ & surface site density [\si{\per\meter\squared}] \\
$Pr$ & Prandtl number \\
$p,p_{H_{2}O}$ & static pressure, partial pressure of water [\si{\pascal}]\\
$\mathbf{Q}$ & source term vector \\
$q_{c}$ & condensation coefficient \\
$R_{s}$ & specific gas constant [\si{\joule\per\kilogram\per\kelvin}]\\
$Re$ & Reynolds number \\
$r,\bar{r},r_{20},r_{crit}$ & radius, mean, surface-averaged, and critical radius [\si{\meter}]\\
$S$ & saturation ratio \\
$Sc_{t}$ & turbulent Schmidt number \\
$T$ & static temperature [\si{\kelvin}]\\
$t$ & time [\si{\second}]\\
$\mathbf{U}$ & conservation vector \\
$u,\mathbf{u}$ & velocity and velocity vector [\si{\meter\per\second}]\\
$V$ & volume [\si{\cubic\meter}]\\
$\mathbf{x}$ & streamline \\
$y,Y$ & mass fraction referenced on component mass and referenced on total mass \\
$y^{+}$ & dimensionless wall distance \\
$\alpha, \beta, \nu$ & parameters according to Young \\
$\alpha_{crit}$ & heat transfer coefficient for a nucleus with $r_{crit}$ [\si{\joule\per\squared\meter\per\second\per\kelvin}]\\
$\gamma$ & isentropic exponent \\
$\Delta{G},\Delta{G_{crit}}$ & difference in Gibbs energy, energy barrier for nucleation [\si{\joule}]\\
$\Delta{T}$ & subcooling [\si{\kelvin}]\\
$\Delta{x_{mixed}},\Delta{x_{unmixed}}$ & geometric length related to engine nozzle [\si{\meter}] \\
$\lambda$ & thermal conductivity [\si{\joule\per\meter\per\second\per\kelvin}]\\
$\mu_{k}$ & statistical moments of size distributions \\
$\nu_{t}$ & eddy viscosity [\si{\meter\squared\per\second}] \\
$\rho$ & density [\si{\kilogram\per\cubic\meter}]\\
$\sigma$ & standard deviation \\
$\sigma_{lg}$ & surface tension [\si{\newton\per\meter}]\\
$\phi$ & non-isothermal correction factor \\
\multicolumn{2}{@{}l}{Subscripts and Superscripts}\\
$g,l,p,s$ & quantity of gas/liquid/particle/solid \\
$H_{2}O$ & quantity of mixture of water phases \\
$hom,het$ & homogenous, heterogeneous \\
$i$ & summation index of dispersed phase \\
$nuc$ & nucleation \\
$sat$ & saturation state \\
$t$ & total quantity \\
$\alpha$ & species phase \\
\end{longtable*}}

\section{Introduction}
\lettrine{M}{odern} propulsion systems must increasingly be evaluated not only for efficiency and emissions, but also for their broader environmental impact. One important contributor to aviation’s non-CO$_2$ climate effects is the formation of condensation trails, or contrails—ice clouds that form when water vapor in aircraft exhaust condenses and freezes under cold atmospheric conditions. Originating within the engine plume, these effects are most likely influenced by fuel composition, thermodynamic conditions, and nozzle design. As aviation moves toward low-carbon fuels and alternative propulsion concepts, accurately predicting and mitigating contrail formation becomes an integral part of propulsion system development.

Contrail cirrus remain one of the largest uncertainties in assessing aviation’s climate impact. Persistent contrails—when formed in ice-supersaturated air masses—can evolve into extended cirrus layers that trap outgoing longwave radiation. On average, the net radiative forcing from contrails is assumed to be warming and, under typical operating conditions, may rival or exceed the effect of aviation’s CO$_2$ emissions~\cite{Burkhardt2011, Schumann2012, Kaercher2018, Lee2021}. Because contrails form within seconds behind the aircraft but evolve over much longer spatial and temporal scales, accurate prediction requires resolving a range of coupled processes—beginning with the near-field formation phase.

This early phase begins as hot, moist exhaust gases mix with cold ambient air at cruise altitude, leading to rapid cooling and dilution, and creating steep thermodynamic gradients and strongly non-equilibrium conditions in a multi-component gas mixture. Water vapor condenses onto soot and other nuclei or directly from the gas phase, forming droplets that may subsequently freeze into ice crystals. These transitions are governed by a complex interplay of mixing, nucleation, growth kinetics, and interphase momentum transfer. Yet despite the inherently non-equilibrium nature of this process, most contrail formation models are based on simplified, equilibrium criteria.

The Schmidt–Appleman criterion (SAC), originally developed by Schmidt~\cite{Schmidt1941} and formalized by Appleman~\cite{Appleman1953}, remains the most widely used approach to estimate contrail onset. It applies bulk thermodynamic thresholds to determine when visible condensation may occur. Schumann~\cite{Schumann1996} reformulated the SAC to suit practical applications and modern thermodynamic data. Extensions such as the Contrail Cirrus Prediction Model~\cite{Schumann2012} enable large-scale simulations of contrail evolution but assume instantaneous condensation and ice formation at the point of saturation. More advanced tools like the Aircraft Plume Chemistry, Emissions, and Microphysics Model~\cite{Fritz2020} incorporate aspects of microphysics and represent expanding plume dynamics. Despite these advancements, most models still rely on thermodynamic thresholds and assume quasi-equilibrium phase transitions, limiting their ability to capture the kinetics of condensation and freezing in the near-field exhaust.

At the same time, theoretical studies have clarified the importance of micro-scale processes in contrail formation. Homogeneous and heterogeneous nucleation depend not only on thermodynamic state, but also on particle properties such as soot morphology, surface energy, and the presence of coatings or chemi-ions~\cite{Kaercher2015, Bier2024}. These effects are especially relevant in the context of alternative fuels. Sustainable Aviation Fuels (SAFs) reduce soot emission and modify surface chemistry, while hydrogen combustion is expected to eliminate soot entirely and increases water vapor emission~\cite{Kaufmann2024}. Empirical studies like the one by Dischl~et~al.~\cite{Dischl2024} confirm that fuel composition has a strong impact on contrail ice crystal formation, which makes accurate modeling of microphysical mechanisms a critical requirement for future contrail prediction tools.

Complementary efforts using large-eddy simulation (LES)~\cite{Unterstrasser2014, Lewellen2020} and coupled wake–microphysics frameworks~\cite{Guignery2012, Vancassel2014} have explored the downstream evolution of contrails, particularly the vortex phase and transition into contrail cirrus. In general, modeling efforts capture plume dynamics and dilution to varying degrees, but primarily focus on ice formation~\cite{Cantin2021}. As a result, the coupling between fluid dynamics, nucleation kinetics, and steep thermodynamic gradients during the early formation phase remains only partially resolved.

A critical yet underrepresented aspect of this early formation phase is the role of the liquid state. While most models emphasize ice crystal number as the key contrail property, likely due to the longer time scales considered, the size distribution of liquid droplets formed during the initial condensation phase directly shapes subsequent freezing behavior. Larger droplets freeze more readily; smaller ones may remain supercooled under certain ambient conditions, leading to a cascade of freezing events as the plume evolves. These interactions between vapor condensation and ice nucleation—although occurring on short timescales—may influence contrail persistence and radiative properties downstream.

In other fields, such as steam turbine modeling, non-equilibrium condensation processes have been extensively studied. These models incorporate detailed thermophysical properties, kinetic relations for nucleation and growth, and polydispersed droplet dynamics~\cite{Young1980, Bakhtar2005}. Adapting such formulations to the aircraft exhaust environment offers a path toward consistent, physics-based modeling of both vapor-to-liquid and liquid-to-ice transitions in the near-field plume. The approach is inherently modular and extensible by design, enabling future studies to incorporate additional physics, alternative fuels, or geometrical configurations without altering core assumptions.

This study presents a computational framework designed to resolve the coupled condensation and freezing processes that initiate contrail formation. Combining methods from steam turbine modeling and cloud microphysics, the model accounts for non-equilibrium phase transitions in compressible, multi-component flows. It is implemented in a parallelized finite-volume solver and applied to a high-bypass turbofan engine exhaust under upper tropospheric conditions. Although simplified in geometry, the setup captures key thermodynamic gradients and enables exploration of fuel and nozzle effects on early-phase contrail behavior. Rather than simulating the full contrail lifecycle or associated radiative effects, the model targets the early formation phase that sets the initial conditions for downstream evolution.

While the present study focuses on computational modeling, it is part of a broader research effort that includes experimental investigations on a test rig replicating contrail formation in a laboratory environment. The combined insights from simulation and experiment will support future model validation and refinement, to be reported in a separate study.

The paper is structured as follows. Section~\ref{sec:Thermophysics} introduces the thermophysical modeling of multi-component mixtures and non-equilibrium phase transitions. Section~\ref{sec:Methodology} presents the governing equations and numerical implementation. In Section~\ref{sec:Results}, results are shown for a nozzle flow replicating aircraft engine exhaust conditions. Section~\ref{sec:Conclusions} summarizes the findings and discusses implications for contrail modeling and sustainable propulsion system design.

\section{Thermophysical Properties}
\label{sec:Thermophysics}
The thermodynamic behavior of multi-component exhaust flows is governed by both the state of the gas mixture and the occurrence of phase transitions. Although these processes are inherently coupled, their numerical treatment benefits from a modular separation: a first subsection focuses on the evaluation of thermodynamic properties in single-phase regions, while the modeling of phase-change mechanisms is addressed in a second subsection.

\subsection{Fluid Properties of Multi-Component Mixtures}

This study introduces a thermodynamic framework for multi-component, non-reacting flows relevant to aircraft exhaust plumes. The modular formulation supports future extensions, including gas-phase chemistry, radiative effects, and additional aerosols. In this initial implementation, species such as sulfuric acid, volatile organic compounds, and other trace aerosols are neglected. This approach isolates the key features of the thermophysical properties while ensuring that the structure remains extensible for later refinement. The methodology follows the rationale adopted in several prior contrail and microphysical models that separate thermochemical effects from microphysics in initial analysis~\cite{Kaercher2003,Lewellen2014}.

The gas phase is modeled using a staggered mixture framework. All non-condensable gas-phase species—such as nitrogen, oxygen, carbon dioxide, argon, and optional minor combustion products—are grouped into a single effective dry-air component. This dry-air surrogate is treated as an ideal mixture of non-ideal gases. Each species is modeled using thermodynamic properties from the REFPROP v10 database~\cite{Lemmon2018}, which allows for density- and temperature-dependent property evaluation without assuming ideal-gas behavior at the component level. Although chemically distinct, these species are grouped into a single pseudo-component to simplify the fluid description, with the option to disaggregate them in future stages if radiative or reactive effects are introduced. As discussed by Schumann~\cite{Schumann1996}, the thermodynamic contribution of carbon dioxide and nitrogen oxides in the early plume region is negligible with respect to the pressure and density fields, although their long-term atmospheric effects are well known.

All thermodynamic properties of water are evaluated using the IAPWS-95 formulation~\cite{Wagner2002} over the full temperature and pressure range of interest. To ensure physically consistent behavior in colder regimes, saturation pressure data below the triple point are obtained using the Sonntag correlation~\cite{Sonntag1994}.
Thermodynamic properties in the metastable regions of vapor and liquid—where a single phase remains locally stable beyond the saturation boundary—are needed for consistent and continuous evaluation near the two-phase region. These states, which lie between the saturation curve and the spinodal (thermodynamic stability limit), are constructed via bilinear extrapolation along the saturation boundary. This provides a smooth extension of all relevant single-phase properties without introducing artificial discontinuities. 

Dispersed phases such as liquid water, ice, and soot are treated thermodynamically as distinct from the gas. Pressure identity is assumed throughout all phases. Soot particles are treated as spherical carbon particles with a polydispersed size distribution specified according to engine conditions. Although soot morphology and surface properties may evolve due to chemical or physical aging, this aspect is not considered at this stage, in line with the simplifications made in laboratory studies and early contrail models~\cite{Moehler2005}.

The thermophysical property evaluation developed in this subsection provides all relevant single-phase quantities across the plume domain, including metastable conditions near saturation. This enables consistent coupling with the non-equilibrium phase-change processes described below.

\subsection{Phase Change Modelling}
\label{subsec:PhaseChangeModelling}
Only water is considered as a condensable species in this study. Phase-change processes involving water vapor—namely condensation and freezing—occur in aircraft exhaust plumes under strongly non-equilibrium conditions, driven by rapid expansion, cooling, and the prevalent water vapor content of engine emissions. Unlike equilibrium models that assume instantaneous phase change at saturation, non-equilibrium models account for energy barriers to nucleation and the finite time scales of droplet and crystal growth. 

Supersaturation provides the thermodynamic driving force for phase change: it arises when the partial pressure of water exceeds the equilibrium saturation pressure at the local gas temperature. This can be expressed either as a dimensionless saturation ratio $S$ or as a subcooling $\Delta T$:

\begin{equation}
    S = \frac{p_{H_{2}O}}{p_{sat}(T)}\qquad\rightleftharpoons\qquad\Delta{T} = T_{sat}(p_{H_{2}O}) - T
    \label{eq:Supersaturation_Subcooling}
\end{equation}

Conditions of $S > 1$ or $\Delta T > 0$~K indicate a local thermodynamic state favorable for phase change, but actual nucleation depends on both thermodynamic thresholds and energetic and kinetic factors.

\begin{figure}[htbp]
    \centering
    \includegraphics[width=1\linewidth]{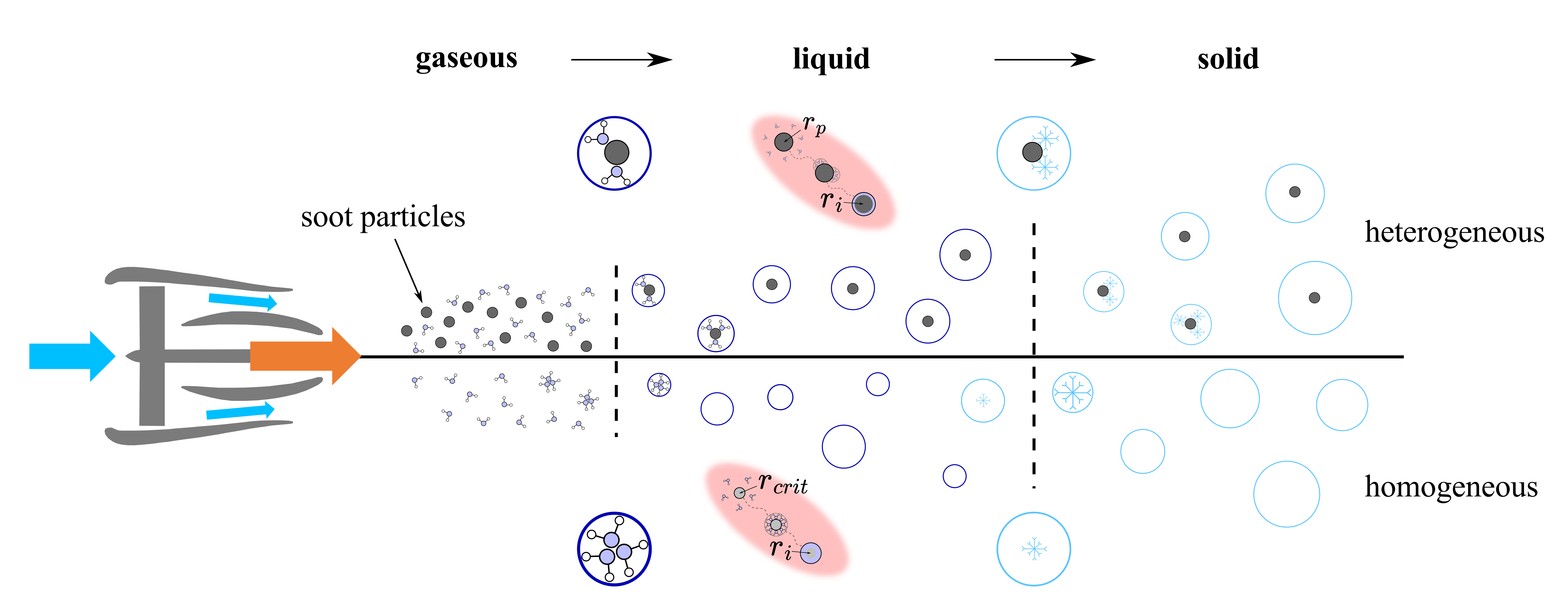}
    \caption{Schematic of the phase change mechanisms involved in contrail formation, inspired by K{\"a}rcher~\cite{Kaercher2018}.}
    \label{fig:Contrail_Formation}
\end{figure}

Figure~\ref{fig:Contrail_Formation} provides a schematic overview of the four main mechanisms relevant to contrail formation. As the jet exhaust mixes with ambient air, condensation can occur either via heterogeneous nucleation on soot particles or homogeneous nucleation directly from the gas phase. At lower temperatures, heterogeneous and homogeneous freezing pathways convert liquid water into ice crystals that make up the visible contrail.

Each of these four mechanisms is governed by distinct thermodynamic and kinetic criteria, yet all originate from the common requirement of local supersaturation. The following sections introduce the models used to describe nucleation and growth, including their coupling with the surrounding gas phase.\\

\subsection*{Homogeneous Condensation}

In the absence of pre-existing condensation nuclei, condensation of water vapor proceeds via homogeneous nucleation, triggered when fluctuations in vapor density produce clusters that exceed a critical size. These clusters are initially unstable and only become thermodynamically favored if the surrounding vapor is sufficiently supersaturated. Classical nucleation theory (CNT)~\cite{Volmer1926, Becker1935} describes this process by evaluating the Gibbs free energy associated with forming a spherical liquid cluster (nucleus) of radius $r$. The energy barrier is given by:

\begin{equation}
    \Delta{G} = \frac{4}{3} \pi r^{3} \rho_{l} (g_{l} - g_{g}) + 4 \pi r^{2} \sigma_{lg}
\end{equation}

The difference in specific Gibbs free energy between liquid and gas is usually simplified to $R_{s} T_{g} \ln{S}$. The critical radius and energy barrier for nucleation are then given by:

\begin{equation}
    r_{crit} = \frac{2\sigma_{lg}}{\rho_{l} R_{s} T_{g} \ln{S}}, \quad
\Delta{G_{crit}^{hom}} = \frac{4}{3} \pi \sigma_{lg} r_{crit}^{2}
\label{eq:Critical_Radius}
\end{equation}

The homogeneous nucleation rate $J^{hom}_{l}$ represents the number of stable nuclei formed per unit volume and time. In its classical form, the rate follows an Arrhenius-type dependence on the energy barrier:

\begin{equation}
    J^{hom}_{l} = q_{c}\sqrt{\frac{2\sigma_{lg}N_{A}^{3}}{\pi{}M^{3}}}\cdot\frac{\rho_{g}^{2}}{\rho_{l}}\cdot\exp\left(-\frac{\Delta{G_{crit}^{hom}}}{k_{b} T_{g}}\right)
\end{equation}

For improved accuracy, especially in non-ideal flows, a correction factor $\phi$ according to Kantrowitz~\cite{Kantrowitz1951} is introduced to account for heat transfer during nucleation:

\begin{equation}
    \phi = q_{c}\frac{\rho_{g}}{\alpha_{crit}}\sqrt{\frac{R_{s}T_{g}}{2\pi}} \left( \frac{h_{fg}^{2}}{R_{s} T_{g}^{2}} - \frac{h_{fg}}{T_{g}} \right)
    \label{eq:Kantrowitz_Correction}
\end{equation}

To track the evolution of the condensed phase, the wetness fraction $y_{l}$ is introduced as the mass fraction of liquid water relative to the total mass of water. The change in $y_{l}$ due to homogeneous nucleation is expressed as:

\begin{equation}
    \left(\frac{dy_{l}}{dt}\right)_{nuc}^{hom} = \frac{4}{3} \pi r_{crit}^{3} \frac{\rho_{l}}{\rho_{H_{2}O}} \cdot \frac{J^{hom}_{l}}{1 + \phi}
\end{equation}

The density of water is expressed based on the wetness fraction and the quantities of the gas and liquid phases:

\begin{equation}
    \rho_{H_{2}O} = (1-y_{l})\rho_{g} + y_{l}\rho_{l}
\end{equation}

The second term, involving the liquid density of the dispersed phase, is evaluated as a finite sum over nucleated droplets distinguished by radius.

\subsection*{Heterogeneous Condensation on Soot Particles}

In the presence of solid surfaces such as soot particles, condensation proceeds via heterogeneous nucleation, a non-equilibrium process that occurs at lower supersaturation levels than required for homogeneous nucleation. The presence of a substrate modifies the interfacial energy balance and reduces the nucleation barrier. Classical nucleation theory accounts for this by introducing a geometric correction that depends on the contact angle $\theta$ between the forming liquid nucleus and the solid surface~\cite{Fletcher1958}. The modified energy barrier is:

\begin{equation}
    \Delta{G_{crit}^{het}} = f(\theta) \cdot \Delta{G_{crit}^{hom}}
\end{equation}

where the reduction factor $f(\theta)$ is given by:

\begin{equation}
    f(\theta) = \frac{1}{4} \left(2 + \cos \theta\right)\left(1 - \cos \theta\right)^{2}
\end{equation}

This formulation assumes a spherical cap geometry for the droplet on the soot surface. Smaller contact angles indicate higher surface wettability, leading to stronger enhancement of the nucleation rate.

Measured contact angles between water and soot under aircraft-relevant conditions range from $59^{\circ}$ for freshly prepared kerosene soot to over $100^{\circ}$ for aged samples~\cite{Popovitcheva2000,Persiantseva2005,Popovicheva2008}. These variations depend on fuel chemistry, combustion conditions, and particle aging. In the present model, a constant contact angle of $90^{\circ}$ is adopted as a representative mid-range value. This simplification is consistent with prior contrail and cloud microphysics studies~\cite{Pruppacher1997} and supports a tractable parametrization of soot activation in the near field.

The critical radius remains unchanged from the homogeneous case in Eq.~\ref{eq:Critical_Radius}, but the nucleation rate is enhanced due to the lowered energy barrier:
\begin{equation}
    J^{het}_{l} = q_{c}\sqrt{\frac{2\sigma_{lg}N_{A}^{3}}{\pi{}M^{3}}}\cdot\frac{\rho_{g}^{2}}{\rho_{l}}\cdot\exp\left( -\frac{f(\theta) \cdot \Delta{G_{crit}^{hom}}}{k_{b} T}\right)
\end{equation}

The contribution to the liquid water mass fraction from heterogeneous nucleation is given by:

\begin{equation}
    \left(\frac{dy_{l}}{dt}\right)_{nuc}^{het} = \frac{4}{3} \pi r_{crit}^3 \frac{\rho_{l}}{\rho_{H_{2}O}} \cdot \frac{J^{het}_{l}}{1 + \phi}
\end{equation}

using the same correction factor $\phi$ as in the homogeneous case in Eq.~\ref{eq:Kantrowitz_Correction}, to account for heat transfer at the droplet surface. Soot particles are modeled as spatially distributed, non-depleting nucleation sites defined at the inflow. Their influence on nucleation is determined solely by the assumed contact angle, which is held constant across the population. While oxidation and aging may alter soot surface properties downstream, the near-field focus of this study justifies neglecting such effects.

\subsection*{Homogeneous Freezing of Aqueous Droplets}

At sufficiently low temperatures, supercooled liquid droplets formed through condensation may undergo homogeneous freezing, leading to the formation of ice particles without requiring external nucleation sites. This process becomes thermodynamically favorable once the water activity inside the droplet falls below a critical, temperature-dependent threshold.

The present formulation adopts the model introduced by Koop et al.~\cite{Koop2000}, where the homogeneous freezing rate $J_{s}^{hom}$ is expressed as a function of the droplet temperature and water activity:

\begin{equation}
    J^{hom}_{s} = f(T_{l}, a_{w})
\end{equation}

Homogeneous freezing acts on the population of liquid-phase droplets formed during prior condensation. The total condensed water is tracked by the wetness fraction $y_{l}$, which is subsequently divided into liquid and solid contributions once freezing begins. Following the approach by Spichtinger and Gierens~\cite{Spichtinger2008}, the evolution of the ice mass fraction due to nucleation is given by:

\begin{equation}
    \left( \frac{dy_s}{dt} \right)_{nuc}^{hom}
    = \frac{1}{\rho_{H_{2}O}} \sum_{i} n_{l,i} \cdot J^{hom}_{s}(T_{l}, a_{w}) \cdot \rho_{s,i} \cdot V_{l,i}^2
\end{equation}

The water density $\rho_{H_{2}O}$ accounts here for all phase contributions and ensures proper normalization of the ice mass fraction. The phase transition from liquid to solid is treated as a direct mass transfer within the condensed phase. Consequently, any increase in the ice mass fraction due to homogeneous freezing results in an equivalent decrease in the liquid water content.

\subsection*{Heterogeneous Freezing - Immersion Freezing}

Of the possible ice formation pathways involving soot, only immersion freezing is included in the present model. Deposition freezing, which requires sub-saturation with respect to liquid water, is unlikely under the warm, moist conditions of the near-field exhaust. Contact freezing, which relies on collisions between supercooled droplets and ambient aerosols, is also excluded due to the absence of background aerosol species in the model~\cite{Kaercher2015}.

Immersion freezing occurs when soot particles are embedded within supercooled droplets, reducing the energy barrier for ice nucleation and enabling freezing at higher temperatures than required for homogeneous processes.

This study adopts the active surface site density approach introduced by Vali~\cite{Vali1971}, in which the immersion freezing rate $J^{het}_s$ is given by:

\begin{equation}
    J^{het}_s = n_{s}(T_{l}) \cdot \sum_{i} n_{p,i} \cdot 4\pi r_{p,i}^2
\end{equation}

For a polydispersed droplet population, the corresponding rate of ice mass formation is given by:

\begin{equation}
    \left( \frac{dy_s}{dt} \right)_{nuc}^{het} = \frac{1}{\rho_{H_2O}} \sum_{i} n_{l,i} \cdot J^{het}_s(T_l) \cdot \rho_{s,i} \cdot V_{l,i}^2
\end{equation}

As in homogeneous freezing, this term represents a redistribution of mass from the liquid to the solid phase. The surface site density $n_s(T)$ characterizes the ice-nucleating efficiency of the soot material. A constant value of $10^{8}$~m$^{-2}$ is used in this study, consistent with experimental data~\cite{Ullrich2017}.

\subsection*{Growth Kinetics}

Once nucleated, droplets grow by vapor deposition. The growth rate $G(r) = dr/dt$ for droplets of radius $r$ is governed by interfacial heat and mass transfer, following the formulation of Young~\cite{Young1980}:
\begin{equation}
    G(r) = \frac{\lambda_{g} \Delta{T} \left( 1 - \frac{r_{crit}}{r} \right)}{\rho_{l} h_{fg} r \left( \frac{1}{1 + 2 \beta Kn} + 3.78(1 - \nu) \frac{Kn}{Pr}\right)}\qquad\text{with}\qquad\nu = \frac{R_{s}T_{sat}}{h_{fg}}\left(\alpha - 0.5 - \frac{2 - q_{c}}{2q_{c}} \cdot \frac{\gamma + 1}{2\gamma} \cdot \frac{c_{p,g}T_{sat}}{h_{fg}}\right)
    \label{eq:Growth_Young}
\end{equation}

The Knudsen number is defined based on the droplet diameter $Kn = \lambda/(2r)$ while the Prandtl number is evaluated using the quantities of the gaseous phase of water. Following common practice for water vapor condensation in low-pressure environments, $\alpha=11$ and $\beta=0$ are used throughout this work. The contribution to the wetness fraction from droplet growth is:
\begin{equation}
    \left(\frac{dy_{l}}{dt}\right)_{growth} = \sum_{i} \frac{n_{l,i}}{m_{H_{2}O}} \cdot 4 \pi r_i^{2} \rho_{l} \cdot G(r_i)
    \label{eq:Mass_Fraction_Growth}
\end{equation}

Ice crystals formed by freezing are assumed to grow analogously to liquid droplets during the early formation phase considered here. The growth rate is evaluated using Eq.~\ref{eq:Growth_Young}, substituting the thermophysical properties of the solid phase. The corresponding contribution to the ice mass fraction follows the same structure as Eq.~\ref{eq:Mass_Fraction_Growth}, with quantities adjusted for ice. Although the resulting growth contributes only modestly to latent heat release over the short timescales of near-field evolution, it is retained in the model for physical completeness and consistency with the treatment of liquid-phase dynamics. 

Combining contributions of both nucleation pathways and growth yields the evolution of the wetness fraction for each phase:
\begin{equation}
    \frac{dy_{l/s}}{dt} = \left(\frac{dy_{l/s}}{dt}\right)_{nuc}^{hom} + \left(\frac{dy_{l/s}}{dt}\right)_{nuc}^{het} + \left(\frac{dy_{l/s}}{dt}\right)_{growth}
    \label{eq:Wetness_evolution}
\end{equation}

This formulation provides a consistent and modular description of phase growth in the early plume, applicable across both droplet and ice crystal populations. Evaporation and sublimation are modeled by reversing the thermodynamic driving potentials of the growth kinetics, providing a consistent and thermodynamically reversible description of phase change dynamics.

\subsection*{Moment-Based Representation of Dispersed Phases}

All dispersed phases—condensed water droplets, ice crystals, and soot particles—are treated as spherical and polydispersed in radius. Due to the time-lag between nucleation and subsequent growth, the droplet and crystal populations naturally evolve toward a continuous size distribution. Similarly, soot emitted from aircraft engines is not monodispersed but follows a characteristic radius distribution determined by combustion processes.

To enable consistent and computationally efficient modeling of these distributions, the present approach follows the method of moments introduced by Hulburt and Katz~\cite{Hulburt1964} and extended by Hill~\cite{Hill1966}. The number density function $f(r)$ describes the distribution of particle radii $r$ in each dispersed phase, and its evolution is governed by nucleation and growth kinetics. The $k$-th moment of the radius distribution is defined as:

\begin{equation}
    \mu_{k} = \int_0^\infty r^{k} f(r) \, dr
\end{equation}

and its partial differential equation with respect to time is given by:

\begin{equation}
    \frac{d\mu_{k}}{dt} = k \int_0^\infty r^{k-1} G(r) f(r) \, dr + J^{hom/het} r_{crit}^{k}
    \label{eq:Method_of_Moments}
\end{equation}

The first term describes the contribution of radius growth to the moment dynamics, while the second term accounts for nucleation. For ice crystal growth, the critical radius $r_{crit}$ is replaced by the actual radius of the supercooled liquid droplet undergoing freezing. To capture the distinct size evolution of droplets formed by heterogeneous versus homogeneous nucleation, a bimodal moment approach is adopted for the liquid phase. Two independent moment sets are evolved separately, allowing the model to track the parallel growth histories of soot-activated and spontaneously nucleated droplets.

To avoid direct numerical integration of the growth term in Eq.~\ref{eq:Method_of_Moments}, a closure approximation is introduced using the surface-area-weighted radius $r_{20}$, following White~\cite{White2003}:

\begin{equation}
    \frac{d\mu_{k}}{dt} = k \mu_{k-1} G(r_{20}) + J^{hom/het} r_{crit}^{k} \quad \text{with} \quad r_{20} = \sqrt{\frac{\mu_2}{\mu_0}}
\end{equation}

This reduces the system to a set of coupled ordinary differential equations for the first few moments, which are solved iteratively in time. The zeroth moment corresponds to the number density, while the third moment can be used to reconstruct the mass fraction of each phase.

The phase-change modeling presented in this section relies on thermodynamically consistent property evaluation across all relevant phases. The accurate treatment of single-phase and metastable states enables physically coherent coupling with nucleation and growth processes. 

\section{Methodology}
\label{sec:Methodology}

The combined treatment of multi-component thermodynamics, non-equilibrium phase-change modeling, and a moment-based description of polydispersed particles forms the basis for simulating contrail formation in expanding aircraft exhaust plumes. This section outlines the governing equations, coupling strategies, and numerical implementation used to resolve thermodynamic and microphysical processes in space and time.

\subsection{Governing Equations}

The compressible flow is modeled using the Reynolds-Averaged Navier–Stokes (RANS) equations in conservative form, extended to multiple interacting phases and species. For each species phase~$\alpha$, the governing equations for mass, momentum, and total energy are:

\begin{equation}
    \frac{\partial \mathbf{U}_{\alpha}}{\partial t} + \nabla \cdot \left( \mathbf{F}^{c}_{\alpha}- \mathbf{F}^{v}_{\alpha} \right) = \mathbf{Q}_{\alpha}
\end{equation}

with the vector of conserved variables defined as:
\begin{equation}
    \mathbf{U}_{\alpha} = (\rho_{\alpha}~~\rho_{\alpha} \mathbf{u}_{\alpha}~~\rho_{\alpha} E_{\alpha})^{\mathsf{T}}, \quad
    E_{\alpha} = e_{\alpha} + \tfrac{1}{2}|\mathbf{u}_{\alpha}|^2, \quad
    H_{\alpha} = h_{\alpha} + \tfrac{1}{2}|\mathbf{u}_{\alpha}|^2
\end{equation}

The source term captures interphase mass, momentum, and energy exchange due to non-equilibrium phase change processes between vapor, liquid, and solid water phases. Dispersed phases—including liquid droplets, ice crystals, and soot particles—are modeled in an Eulerian frame of reference to ensure numerical stability and scalability. Separate RANS equations are solved for the dry-air mixture and all three phases of water. Interphase momentum exchange is modeled via a drag force source term based on the Schiller–Naumann correlation with a Cunningham slip correction:

\begin{equation}
    C_D = \frac{24}{Re} \left(1 + 0.15 Re^{0.687} \right) \cdot \frac{1}{1 + 2A_C Kn}\qquad\text{with}\qquad
    A_{C} = 1.257 + 0.4 \exp\left(\frac{-1.1}{2\text{Kn}}\right)
\end{equation}

where Reynolds and Knudsen numbers are based on radius. The Stokes number is used to quantify slip effects. Scalar transport equations are solved for selected gas-phase species, notably water vapor and the dry-air mixture (modeled as a pseudo-component). The mass fraction~$Y_j$ of species~$j$ evolves according to:

\begin{equation}
    \frac{\partial (\rho Y_j)}{\partial t} + \nabla \cdot (\rho Y_j \mathbf{u}) = \nabla \cdot \left( \rho D_j \nabla Y_j \right) - \rho \left(\frac{dy_{j,l}}{dt}+\frac{dy_{j,s}}{dt}\right)
\end{equation}

This formulation ensures mass conservation across phases. Source terms for water vapor are derived from the non-equilibrium phase-change model in Section~\ref{subsec:PhaseChangeModelling}. Dry air has no associated sources or sinks, and soot is passively transported. Molecular diffusion of water vapor is modeled based on a diffusion coefficient scaling approximately as $D \propto T^{1.94}/p $, as discussed in, e.g.,~\cite{Pruppacher1997}.

Turbulent diffusion is modeled using the gradient diffusion hypothesis. A constant turbulent Schmidt number of~$Sc_{t} = 0.7$ is assumed, such that $D_j = \nu_t / Sc_t$, where $\nu_t$ is provided by the Spalart–Allmaras turbulence model.

In addition, moment transport equations are solved for the radius distributions of each dispersed phase:

\begin{equation}
    \frac{\partial (\rho \mu_k)}{\partial t} + \nabla \cdot (\rho \mu_k \mathbf{u}) = \nabla \cdot \left( \rho D_k \nabla \mu_k \right) + \rho \frac{d\mu_k}{dt}
\end{equation}

The source term $d\mu_k/dt$ accounts for nucleation and growth and is defined in Section~\ref{subsec:PhaseChangeModelling}. Soot particles are assumed non-reactive and non-condensing, hence no source terms are included in the corresponding set of equations.

Dispersed phases evolve independently, and interactions such as coagulation or collision are not considered. However, condensation and freezing processes are modeled as interphase mass transfer, with consistent coupling between vapor, liquid, and solid water through shared source terms. Radiative heat transfer is neglected, in line with typical assumptions for near-field plume modeling~\cite{Lewellen2014,Unterstrasser2017}.

\subsection{Numerical Methods}

The simulations are conducted using a structured-grid, density-based compressible flow solver developed in-house and extended to support multi-component, multi-phase flows with non-ideal thermodynamics~\cite{Tegethoff2024}. The solver is applied in a steady-state formulation and solves the Reynolds-Averaged Navier–Stokes (RANS) formulations in two or three spatial dimensions.

Convective fluxes are computed using a second-order Phase Generalised Ideal Roe (PGIRoe) scheme~\cite{Tegethoff2025}, which provides enhanced robustness in transcritical and phase-changing flow regimes. A MUSCL-type spatial reconstruction with the van Leer limiter is applied for second-order accuracy, while viscous fluxes are discretized using central differences.

A pseudo-time-stepping strategy is used to obtain steady-state solutions, employing a second-order explicit Runge–Kutta method with a globally controlled CFL number. Non-equilibrium source terms due to condensation and freezing are evaluated explicitly in each cell at every subiteration and integrated directly into the conservative source term vector, ensuring tight coupling between the microphysics and flow field evolution. The dispersed phases are modeled in an Eulerian frame of reference. The solver architecture also supports coupling to a Lagrangian particle-tracking framework for the dispersed phases, enabling hybrid Eulerian–Lagrangian approaches if desired.

Turbulence closure is provided by the Spalart–Allmaras model, offering a robust balance between physical accuracy and numerical stability. The REFPROP v10 library serves as the thermophysical backend, enabling consistent evaluation of non-ideal fluid properties across the relevant temperature and pressure ranges. Thermodynamic and transport properties of all phases of water are tabulated prior to runtime for improved computational efficiency. During runtime a second-order Taylor series expansion is employed for interpolation of tabulated thermophysical properties.

The solver includes parallel backends for both CPU and GPU architectures. The CPU implementation uses MPI-based domain decomposition for distributed-memory parallelism. GPU acceleration is supported via a CUDA-based GPU backend with structured-grid support and explicit time integration. Core kernels are ported using a structure-of-arrays layout, and the implementation scales efficiently across GPUs via MPI domain decomposition.

\section{Results}
\label{sec:Results}
\subsection{Geometry and Flow Setup of Aircraft Nozzle}

\begin{figure}[b!]
    \centering
    \includegraphics[width=1\linewidth]{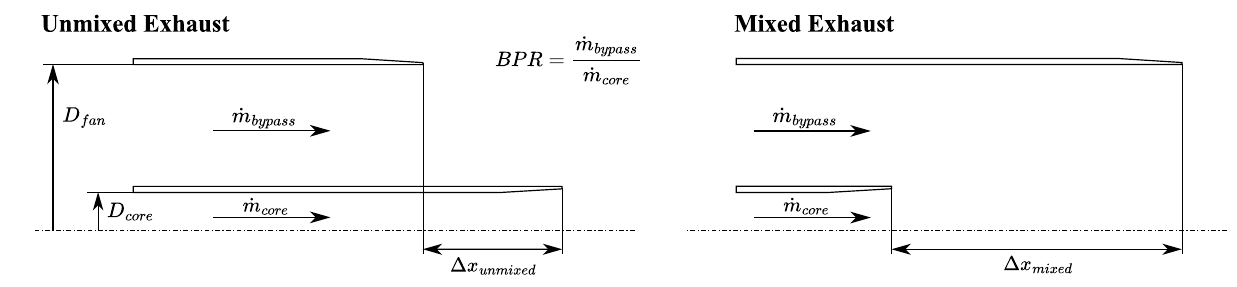}
    \caption{Schematic of the unmixed and mixed nozzle configurations used in this study.}
    \label{fig:Nozzle_Schematic}
\end{figure}

The simulations presented in this study are based on a simplified axisymmetric representation of a high-bypass turbofan engine under cruise conditions. Two nozzle configurations are investigated: an unmixed exhaust scenario in which the core and bypass streams remain segregated until leaving the nozzle, and a mixed exhaust case in which both streams are fully combined prior to exit. The nozzle geometry is parameterized using the engine's bypass ratio (BPR), with the fan diameter $D_{fan}$ serving as the primary reference length. The core nozzle diameter $D_{core}$ is determined based on the area ratio implied by the bypass ratio as
$D_{core} = \sqrt{\frac{D_{fan}^{2}}{1 + {BPR}}}$.
The axial length of the unmixed configuration is approximated as $\Delta{x_{unmixed}} = 0.4D_{fan}$, representing a short co-annular section prior to exhaust into the ambient. In contrast, the mixed configuration uses an effective downstream section of $\Delta{x_{mixed}} = 3 D_{core}$, corresponding to a nozzle representative of mixed-flow designs. A summary of the geometric parameters used in this study is provided in Table~\ref{tab:Nozzle_Geometry}, based on public-domain data for the Rolls-Royce Trent~1000~\cite{RollsRoyce_Trent1000}. Both geometry configurations are illustrated schematically in Fig.~\ref{fig:Nozzle_Schematic}.

\begin{table}[b!]
\centering
\caption{Nozzle geometry parameters for the Rolls-Royce Trent~1000~\cite{RollsRoyce_Trent1000}.}
\label{tab:Nozzle_Geometry}
\begin{tabular}{lcc}
\hline
\textbf{Parameter} & \textbf{Symbol} & \textbf{Value} \\
\hline
Fan diameter        & $D_{fan}$     & 2.84 m \\
Bypass ratio        & $BPR$                  & 10 \\
Core diameter       & $D_{core}$    & 0.86 m \\
Unmixed section length & $\Delta{x_{unmixed}}$ & 1.14 m \\
Mixed section length   & $\Delta{x_{mixed}}$   & 2.58 m \\
\hline
\end{tabular}
\end{table}

The computational domain extends significantly beyond the nozzle exit to ensure undisturbed entrainment and realistic jet development. The axial length of the domain is set to $L_x = 8 D_{fan}$, and the radial extent is $L_r = 3 D_{fan}$. This enlarged domain ensures that far-field boundary conditions do not interfere with the evolution of the exhaust plume or the onset of condensation. Non-reflecting boundary conditions are applied at all external boundaries, and ambient properties (temperature, pressure, humidity) are prescribed to match cruise-level atmospheric conditions. These domain dimensions were confirmed to be sufficient through sensitivity tests, showing no appreciable impact on the flow field or thermodynamic state near the nozzle exit.

A structured mesh is used for all simulations, with local refinement near solid surfaces and in the shear layer region. The boundary layers are fully resolved, with a first-cell spacing selected to maintain $y^+ < 1$ across all solid boundaries. The mesh contains approximately two million cells. This resolution exceeds what is typically required for aerodynamic accuracy, but was deliberately chosen to ensure adequate resolution of thermophysical gradients and phase changes. Grid convergence was verified by applying the procedure documented by Roache~\cite{Roache1997}, confirming mesh-independent results for key quantities such as static pressure, mass fraction of water vapor, and droplet size distribution. Additionally, convergence with respect to the resolution of the thermophysical property tables was verified through iterative refinement, ensuring that interpolation yields thermodynamically consistent results as the resolution approaches infinity.

Nozzle walls are treated as adiabatic, with no-slip conditions applied to the gas phases. For the dispersed phases, a purely reflective boundary condition is assumed, consistent with the dilute limit and the short residence time of dispersed phases near the wall. Although three-dimensional simulations were conducted to verify general flow behavior, the results presented here are based on two-dimensional axisymmetric computations to enable detailed analysis of nucleation onset and growth kinetics.

\begin{table}[b!]
\centering
\caption{Boundary conditions applied to the computational domain.}
\label{tab:BC_Conditions}
\begin{tabular}{lcccc}
\hline
\textbf{Property} & \textbf{Core} & \textbf{Bypass} & \textbf{Freestream} & \textbf{Farfield} \\
\hline
Total temperature $T_t$ [K]        & 640   & 281  & -- & --\\
Total pressure $p_t$ [bar]         & 0.40  & 0.53 & -- & --\\
Static temperature $T$ [K]         & --   & --  & 216.5 & 216.5\\
Static pressure $p$ [bar]          & --  & -- & 0.2263 & 0.2263\\
Water vapor mass fraction $Y_{H_2O}$ & 0.025 & 0    & 0  & 0\\
Mach number                        & -- & -- & 0.8 & --\\
Composition                        & dry air + H$_2$O + soot & dry air & dry air & dry air \\
Soot size distribution             & log-normal, $\bar{r}=50$~nm, $\sigma=30\%$ & -- & -- & -- \\
\hline
\end{tabular}
\end{table}

Boundary conditions are defined to replicate cruise-level operation at 11~km altitude and a flight Mach number of $Ma = 0.8$, based on the International Standard Atmosphere (ISA). The domain includes three inflow regions—core, bypass, and freestream—and far-field boundaries. Core and bypass inlets are prescribed using total temperature, total pressure, and gas composition. Core conditions are estimated based on publicly available Trent~1000 engine data at cruise, using reported pressure ratios and component efficiencies in combination with standard thermodynamic relations. Assuming kerosene combustion, the resulting core flow conditions are $T_{t} = 640$~K, $p_{t} = 0.40$~bar, and a water vapor mass fraction of 2.5\%. The core stream includes soot typical for kerosene and modeled via a log-normal distribution ($\bar{r}_{p} = 50$~nm, $\sigma = 30\%$) with an assumed number concentration of $10^8~\mathrm{cm}^{-3}$. The bypass stream corresponds to dry ambient air compressed by the fan, with a pressure ratio of 1.54, yielding $T_{t} = 281$~K and $p_{t} = 0.53$~bar.

The freestream inlet surrounds the nozzle laterally and is defined by static pressure, static temperature, and velocity to represent undisturbed cruise-level flow. ISA values of $T = 216.5$~K and $p = 0.2263$~bar are applied, with a Mach number of 0.8. Unlike the engine inlets, the freestream is not derived from a stagnation process and is specified in static form. Far-field boundaries use the same ISA static conditions and are located sufficiently far from the nozzle to prevent interference with plume development. Dry air is assumed for the bypass and freestream regions, with a molar composition of 78\% N\textsubscript{2}, 21\% O\textsubscript{2}, and 1\% Ar/CO\textsubscript{2}. Based on freestream properties and the fan diameter as the characteristic length scale, the flow corresponds to a Reynolds number of $Re\approx1.7\times10^{7}$, indicating fully turbulent jet conditions. A summary of all boundary conditions is given in Table~\ref{tab:BC_Conditions}.

\subsection{Influence of Non-Equilibrium Phase Changes on Contrail Formation}

Based on the prescribed inflow and ambient conditions, phase-change behavior in the near-field exhaust is analyzed to assess the thermodynamic drivers of condensation and freezing under non-equilibrium conditions. To evaluate the potential for phase transition, the thermodynamic state of the exhaust is first analyzed in the absence of any phase-change modeling. This configuration allows assessment of the saturation environment resulting purely from flow expansion and mixing processes, without feedback from latent heat release or the presence of dispersed phases other than soot particles.

\begin{figure}[b!]
    \centering
    \includegraphics[width=0.99\linewidth]{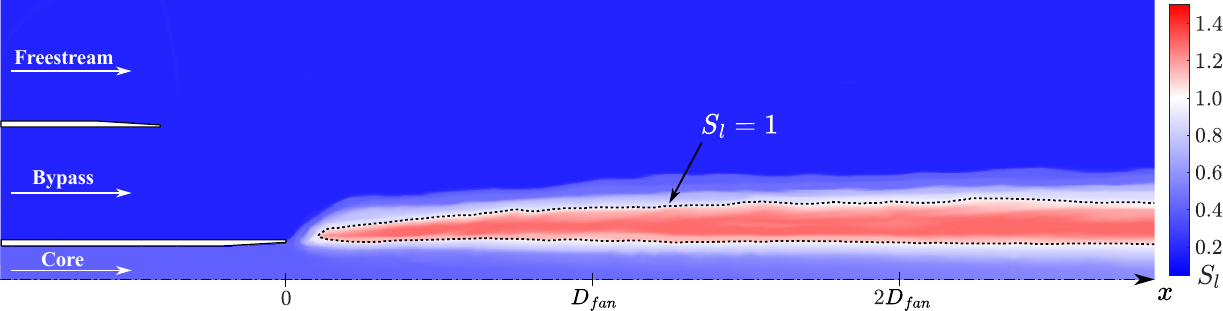}
    \caption{Saturation ratio over liquid water $S_{l}$ in the unmixed nozzle configuration without phase change.}
    \label{fig:Saturation_Ratio}
\end{figure}

Figure~\ref{fig:Saturation_Ratio} shows the spatial distribution of the saturation ratio with respect to liquid water defined in Eq.~\ref{eq:Supersaturation_Subcooling} in the unmixed nozzle configuration. Upon nozzle exit, the hot core stream mixes with the cooler bypass and ambient air while also expanding due to the pressure gradient. This combined effect leads to a strong reduction in static temperature and initiates supersaturation in regions where water vapor concentrations remain sufficiently high. As a result, the water vapor partial pressure can exceed the saturation vapor pressure over liquid water. Although the ambient temperature is well below the freezing point, the saturation ratio with respect to liquid is used here as the liquid phase acts as a necessary intermediate in the pathway to ice.

A supersaturated region with $S_l > 1$ forms in the mixing layer downstream of the core nozzle, centered around a streamline that runs approximately parallel to the centerline but located near the height of the core nozzle, within the developing shear layer. This region emerges closely behind the nozzle exit and extends several fan diameters downstream. The supersaturation gradually decays in both magnitude and spatial extent due to turbulent mixing with the surrounding cooler and drier bypass and freestream air. The central core region remains subsaturated due to elevated temperatures, while the outer freestream remains too dry to support appreciable water vapor content. Consequently, supersaturation develops preferentially in an annular layer where both temperature and water vapor concentrations fall within an intermediate range. The position of the $S_{l} = 1$ isocontour, indicated by a dotted line in Fig.~\ref{fig:Saturation_Ratio}, demarcates the threshold above which condensation would be thermodynamically favorable in the absence of kinetic limitations.

This case forms the baseline for evaluating the role of equilibrium and non-equilibrium phase-change modeling. It confirms that significant condensation potential exists solely due to the underlying thermodynamic evolution of the exhaust plume, independent of any feedback mechanisms of phase change.

\subsubsection{Modeling Sensitivities: Equilibrium vs. Non-Equilibrium Phase Change}

Consistent with the preceding analysis of saturation in the absence of phase change, the following comparison focuses on the near-field region within one fan diameter downstream of the nozzle exit. This interval encompasses the initial region of thermodynamic supersaturation and includes the earliest possible onset of condensation. 

The influence of phase-change modeling on the formation of liquid and ice phases in aircraft exhaust is assessed by comparing two physically distinct representations: an equilibrium approach, which assumes instantaneous phase change once saturation is reached, and a non-equilibrium model, which accounts for delayed condensation due to kinetic limitations associated with nucleation and droplet growth. Both models are evaluated using an identical thermodynamic setup and nozzle geometry, allowing for a direct assessment of the impact of phase-change modeling on the resulting flow features.

\begin{figure}[t!]
    \centering
    \includegraphics[width=1\linewidth]{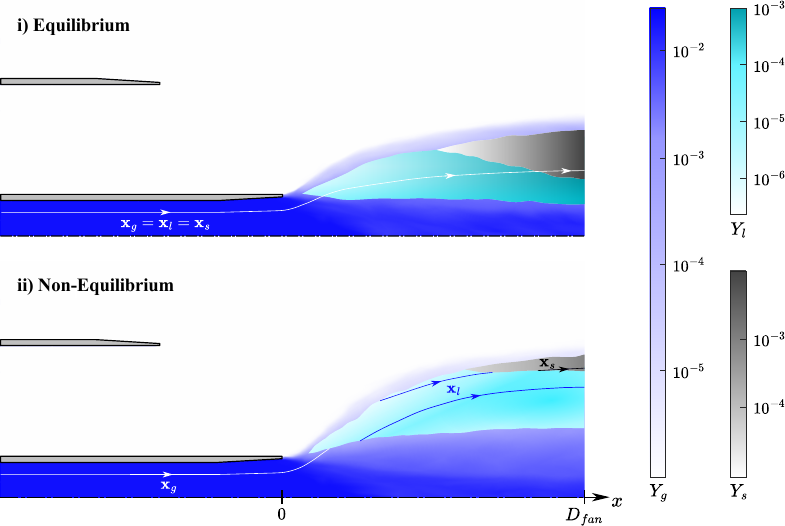}
    \caption{Comparison of phase change models in the unmixed nozzle configuration. i)~Equilibrium assumption. ii)~Non-equilibrium phase changes. Background contours show mass fraction of water vapor, overlaid with mass fraction contours of liquid water (cyan) and ice (gray). Lines represent stream lines.}
    \label{fig:Phase_Change_Models}
\end{figure}

Figure~\ref{fig:Phase_Change_Models} presents the mass fraction of water vapor as a background contour, with overlaid mass fraction contours of liquid water and ice shown in cyan and gray, respectively. Streamlines for each phase are included, illustrating how vapor, liquid, and ice components are transported differently, particularly under non-equilibrium conditions. The results highlight clear differences in the onset, spatial distribution, and magnitude of both condensation and freezing, governed solely by the underlying phase-change representation.

In the equilibrium case, condensation begins immediately once the local saturation ratio exceeds unity. Water vapor is assumed to transition instantaneously to liquid, without accounting for the energy barrier associated with nucleation. As a result, liquid droplet formation occurs near the nozzle exit, in a region similar in shape to that identified in Fig.~\ref{fig:Saturation_Ratio} for the case without phase change. The contours show rapid accumulation of liquid mass within the first fan diameter downstream of the core nozzle, reaching peak values of approximately $9 \times 10^{-4}$. The liquid region closely follows saturation, with higher vapor mass fraction near the soot-laden core and significantly lower values in the outer mixing region, where dilution reduces vapor content. As the local conditions reach saturation with respect to ice, freezing sets in. The ice phase quickly envelops the liquid core, forming a broad and continuous frozen region. Within one fan diameter downstream, the ice mass fraction reaches about $3 \times 10^{-3}$. With increasing distance, the ice region expands and increasingly dominates, displacing the liquid phase. Under the equilibrium assumption, all phases follow the same streamline, along which phase transitions occur consecutively.

In contrast, the non-equilibrium model accounts for finite rates of nucleation and growth. Condensation does not occur immediately at saturation but is delayed until sufficient thermodynamic and kinetic conditions are met. As shown in the lower panel of Fig.~\ref{fig:Phase_Change_Models}, the liquid region is narrower, shifted slightly farther downstream, and displaced radially outward. This reflects the need for both time and thermodynamic favorable conditions for nucleation. 
The maximum liquid water mass fraction is significantly lower than in the equilibrium case, indicating that kinetic effects not only delay condensation but also limit the total condensed mass. Ice formation is likewise delayed and spatially confined. It appears only after sufficient liquid has accumulated and cooled, and remains more fragmented compared to the continuous distribution in the equilibrium model, both in magnitude and coherence. Ice primarily forms in the outer jet region, where mixing and cooling are more favorable.

A parcel of water vapor exiting the core nozzle is immediately exposed to strong radial gradients in temperature and water vapor concentration, as the surrounding bypass stream is substantially cooler and dry. The resulting diffusive fluxes drive vapor radially outward, while mixing-induced cooling pushes the parcel toward saturation. Under non-equilibrium conditions, local supersaturation can be sustained before nucleation occurs, allowing the vapor to spread significantly before phase change regulates the thermodynamic state. Depending on the nucleation mechanism, heterogeneous nucleation near the core region or homogeneous nucleation farther out can eventually trigger droplet formation. Once condensation begins, the release of latent heat and subsequent vapor depletion modify the local thermodynamic gradients, enhancing small-scale mixing. Importantly, because vapor, liquid, and ice phases are treated as separate momentum carriers, differential transport leads to distinct phase streamlines, as visible in Fig.~\ref{fig:Phase_Change_Models}. Vapor exhibits a broader spreading angle compared to the condensed phases, and the formation regions of liquid water and ice are spatially separated from the core outwards. This phase separation is a direct consequence of the finite kinetics of nucleation and growth, and is absent in the equilibrium model.

Beyond these local differences in phase composition, the overall shape of the predicted contrail differs substantially. The equilibrium model produces a broad, symmetric contrail structure with continuous ice coverage around the liquid core. The non-equilibrium case, by contrast, yields a narrower, more asymmetric structure, with ice formation concentrated along the outer edges of the plume. These differences arise from delayed nucleation, localized condensation, and limited growth rates, underscoring the sensitivity of early contrail morphology to the phase-change formulation.

These findings have important implications for modeling contrail microphysics. While equilibrium models may be useful for identifying regions of thermodynamic feasibility, they tend to overpredict condensed and frozen mass and the spatial extent of the contrail. Non-equilibrium models provide a more physically consistent prediction of the timing, location, and intensity of phase change by accounting for kinetic constraints.

In summary, the equilibrium model provides an upper bound on formation of liquid and ice, while the non-equilibrium model yields a more conservative and physically grounded estimate. The latter also enables identification of locally dominant phase-change mechanisms and captures significant differences in contrail shape and extent. Although the present model does not resolve long-term plume dynamics, such early differences may influence downstream evolution, highlighting the importance of accurate near-field modeling in assessments of contrail impact. It should be noted that in all cases considered, the simulated ice crystal concentrations and sizes remained below typical optical visibility thresholds for contrails~\cite{Kaercher2009}.

\subsubsection{Microphysical Evolution of the Liquid Phase}

To further characterize the effects of non-equilibrium condensation, this section examines the local nucleation mechanisms and the resulting droplet size spectrum in the early contrail region. While most contrail models emphasize ice crystal number as the primary microphysical quantity, the present analysis focuses on the liquid phase to resolve the kinetic onset of condensation and the subsequent droplet growth history. This distinction is essential, as the distribution of liquid droplet sizes plays a central role in governing freezing behavior. Although these processes occur over short timescales, they influence the spatial and temporal development of contrails and may have downstream implications.

\begin{figure}[b!]
    \centering
    \includegraphics[width=1\linewidth]{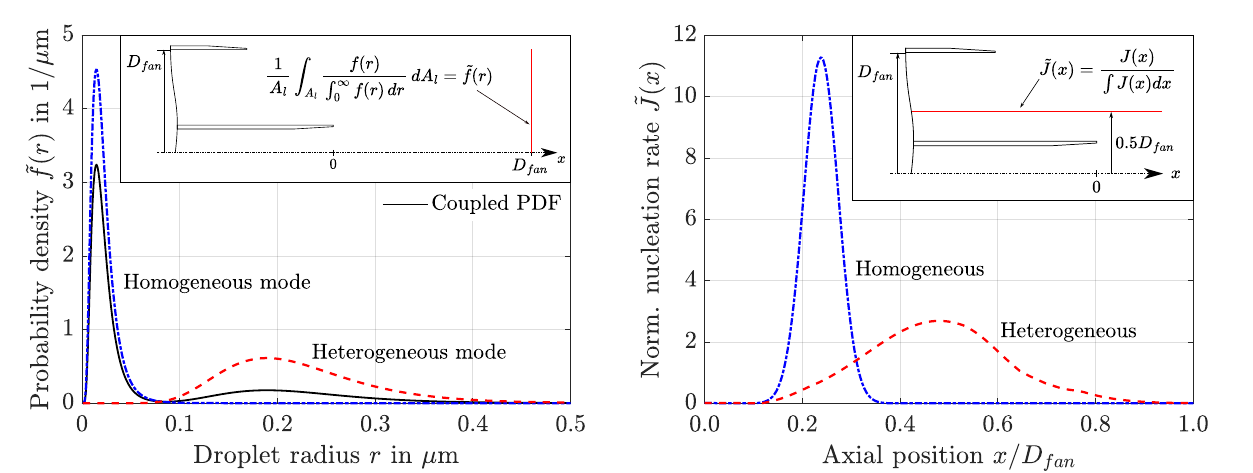}
    \caption{Left: Probability density function of liquid droplets at $x=D_{fan}$ in the non-equilibrium case, resolved by nucleation mechanism. Right: Normalized homogeneous and heterogeneous nucleation rates along a line at height of $0.5D_{fan}$.}
    \label{fig:Droplet_Distribution}
\end{figure}

The left part of Fig.~\ref{fig:Droplet_Distribution} shows the probability density function of droplet radius, area-averaged over the pseudo-surface of the liquid phase at an axial location of $x = D_{fan}$, representative of the early post-nucleation region. The resulting size distribution exhibits a bimodal structure. A sharp peak centered at approximately $r \approx 0.015~\mu\mathrm{m}$ is attributed to homogeneous nucleation, while a broader secondary mode near $r \approx 0.2~\mu\mathrm{m}$ arises from soot-activated heterogeneous condensation. The relative prominence of these peaks reflects the nucleation activity of each mechanism under the prevailing thermodynamic conditions. Notably, while homogeneous nucleation dominates in terms of droplet number, heterogeneous nucleation contributes significantly to the total condensed mass due to larger droplet radii. This distinction is important for predicting freezing onset and crystal growth dynamics, as the size of liquid precursors strongly affects ice nucleation pathways.

To further elucidate the spatial structure of nucleation mechanisms, the right part of Fig.~\ref{fig:Droplet_Distribution} presents the normalized homogeneous and heterogeneous nucleation rates along a horizontal line at a height of $0.5 D_{fan}$. As can be analyzed from Fig.~\ref{fig:Phase_Change_Models}, this line intersects the liquid core horizontally. The profiles indicate that homogeneous nucleation peaks slightly upstream of heterogeneous nucleation. As a result, the side of the jet facing the bypass stream is dominated by homogeneously nucleated droplets, whereas the core-side region is shaped primarily by heterogeneous nucleation.

At first glance, the dominance of homogeneous nucleation near the bypass side appears counterintuitive, given the earlier discussion of saturation levels. However, in contrast to the case without phase change, non-equilibrium effects lead to localized increases in liquid supersaturation due to enhanced mixing and thermal gradients enabling distinct local regions of homogeneous nucleation.

The magnitude difference between the two nucleation rates aligns with expectations based on the difference in activation energy barriers. In regions where both mechanisms are active, heterogeneous nucleation ultimately dominates due to its lower critical energy threshold. Nonetheless, the zone of homogeneous nucleation is spatially narrow but exhibits an explosive nucleation response, resulting in a large number of very small droplets. This effect explains why the homogeneous mode dominates the number-based size distribution despite being confined to a limited spatial region.

Taken together, these results highlight how finite-rate effects shape the initial droplet spectrum and establish the liquid phase as more than a transient intermediate. Rather, it plays a defining role in early plume evolution, determining when, where, and how ice formation may occur. The distribution and evolution of liquid droplets seem therefore critical to understanding and accurately modeling contrail microphysics.

\subsection{Parametric Study: Effect of Fuel Type and Nozzle Geometry}

To explore the sensitivity of contrail formation to propulsion system parameters, a simplified parametric study varies both the fuel type—approximated through changes in water vapor mass fraction—and the nozzle geometry. While the thermodynamic and geometric setup remain identical to the baseline configuration described in previous sections, the water vapor content is increased from 2.5\% to 20\% by mass to approximate hydrogen combustion conditions. In addition, soot is excluded from the simulation, consistent with the near-zero soot emissions expected from hydrogen-fueled engines. Although the overall engine design would likely differ in a hydrogen-based propulsion system, this low-effort approach provides a first-order estimate of microphysical behavior in the absence of detailed hydrogen engine data, which is currently not available in the literature. The chosen value of 20\% water vapor is based on order-of-magnitude estimates from Bier et al.~\cite{Bier2024} and reflects the expected increase in water vapor emission for the same released combustion energy.

\begin{figure}[t!]
    \centering
    \includegraphics[width=1\linewidth]{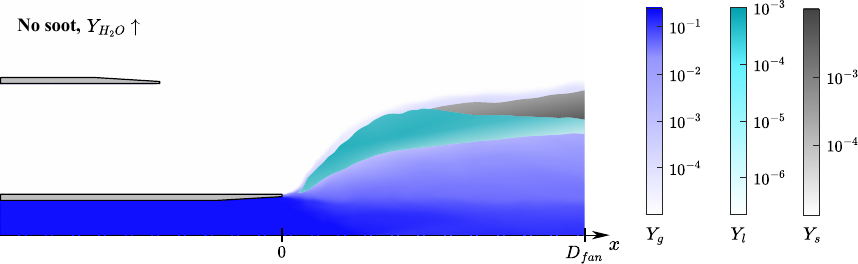}
    \caption{Early-stage phase change in the unmixed nozzle configuration with increased water vapor content and no soot. Background contours show mass fraction of water vapor, overlaid with mass fraction contours of liquid water (cyan) and ice (gray).}
    \label{fig:Soot_Comparison}
\end{figure}

Figure~\ref{fig:Soot_Comparison} presents the simulation results using the same visualization layout as in the earlier comparison of phase-change models. This enables a direct comparison of vapor, liquid, and ice mass fractions between the kerosene-based baseline and the hydrogen-like configuration. The non-equilibrium phase-change model was retained in both cases.

Three key observations emerge from this comparison.
First, the vapor and liquid fields exhibit a broader radial spread compared to the baseline case. This reflects the effect of the increased water vapor mass fraction, which enhances local supersaturation during mixing and allows vapor to extend farther outward before condensation regulates the thermodynamic state. In the absence of soot particles, homogeneous nucleation dominates, and the liquid and ice phases occupy a wider region across the plume.

Second, the liquid phase distribution changes significantly. Without soot particles, only homogeneous nucleation can initiate droplet formation. As a result, the liquid region exhibits locally steeper gradients in mass fraction and a narrower core structure. The increased vapor availability, combined with the lack of competition from heterogeneous nucleation sites, allows homogeneously nucleated droplets to form and grow rapidly. This leads to a dense population of smaller droplets formed through homogeneous nucleation.

Third, the ice phase appears earlier and spreads more aggressively into the liquid region. Freezing begins as early as within the first half fan diameter downstream of the nozzle, particularly in the outer regions of the mixing layer. The broader ice coverage and earlier transition from liquid to ice suggest that homogeneous nucleation alone may be sufficient to initiate contrail formation under hydrogen combustion conditions. While the absence of soot particles would typically reduce ice nucleation probability, the early formation of numerous small liquid droplets provides an alternative freezing pathway. It should be noted, however, that other ambient aerosol particles, not accounted for in the present model, could also act as heterogeneous ice nucleation sites and influence the freezing behavior. Nevertheless, the rapid onset of freezing observed here may enhance droplet growth kinetics and strengthen interactions between the liquid and ice phases.

Although the long-term evolution of the contrail is beyond the scope of the current model, these results illustrate that initial microphysical behavior differs markedly between conventional and hydrogen-like combustion. In particular, the findings suggest that hydrogen-fueled aircraft may generate contrails through homogeneous pathways alone, potentially altering droplet size spectra and phase partitioning. These early differences may have downstream implications and highlight the importance of continued research on hydrogen combustion microphysics.

\begin{figure}[t!]
    \centering
    \includegraphics[width=1\linewidth]{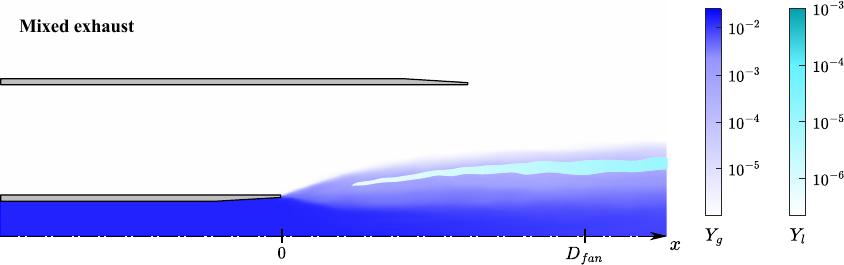}
    \caption{Early-stage phase change in the mixed nozzle configuration. Background contours show mass fraction of water vapor, overlaid with mass fraction contours of liquid water (cyan).}
    \label{fig:Geometry_Comparison}
\end{figure}

To assess the influence of internal mixing on early-phase change dynamics—including both condensation and freezing—a simulation was performed using the mixed nozzle configuration introduced in Fig.~\ref{fig:Nozzle_Schematic}, applying the same non-equilibrium phase-change model as before. All thermodynamic, chemical, and numerical parameters were held constant, allowing a direct comparison focused solely on the impact of geometric variation on the evolution of the condensed phases. Figure~\ref{fig:Geometry_Comparison} presents the resulting field of water vapor mass fraction, overlaid with contours of liquid water mass fractions, using the same visualization approach as in previous figures to ensure comparability.

The modified geometry leads to enhanced internal mixing prior to nozzle exit, producing a more homogenized exhaust with smoother gradients and reduced local vapor content. As a consequence, both condensation and freezing are delayed. Droplet formation begins farther downstream than in the unmixed case, and no ice is detected within the region shown. The liquid phase is spatially narrower, and its peak mass fraction is significantly reduced. While heterogeneous nucleation remains the dominant pathway, the suppressed local supersaturation prevents extensive condensate formation or rapid growth. These findings indicate the role of nozzle mixing strategy in controlling the early thermodynamic conditions that govern condensation onset and freezing kinetics. Although the current analysis is limited to steady, axisymmetric conditions and a simplified geometry, the results demonstrate that internal mixing not only may alter the likelihood of phase change, but also affects the timing, spatial development, and relative partitioning between liquid and ice. Such differences are critical for setting the initial conditions of ice crystal populations in contrail evolution models and underscore the importance of accounting for nozzle design in contrail impact assessments.

\section{Conclusions}
\label{sec:Conclusions}

This study presents a physics-based numerical framework for modeling early-stage contrail formation in the near-field exhaust of aircraft engines. Building on established methods from steam turbine condensation modeling and cloud microphysics, the approach resolves non-equilibrium phase transitions in compressible, multi-component flows. The model incorporates homogeneous and heterogeneous nucleation, interphase momentum exchange, and polydispersed tracking of liquid droplets, ice crystals, and soot particles.

Results for an engine nozzle flow under cruise conditions highlight the significant impact of non-equilibrium effects on both condensation and freezing behavior. Equilibrium-based models, by contrast, tend to overpredict the amount and spatial extent of condensed phases. The simulations reveal that nozzle geometry modulates local thermodynamic conditions, affecting both the onset and intensity of phase change. In particular, unmixed configurations—where bypass and core streams exit separately—produce sharper gradients and earlier condensation and freezing than mixed-flow designs. The mixed exhaust configuration further highlights how the local thermodynamic state, rather than just overall conditions, controls the initial onset of phase transitions. 

Similarly, a parametric study approximating hydrogen combustion conditions through increased water vapor mass fraction and soot exclusion indicates that contrail formation may still occur via homogeneous nucleation alone. These results suggest that early-phase microphysics are highly sensitive to propulsion parameters, even when geometric or thermodynamic variations are modest.

The overarching aim of this study is not to model the full lifecycle of contrail evolution, but to provide a simplified, yet physically consistent, parameter study that highlights the importance of capturing all relevant phase transitions and dispersed-phase dynamics. The findings reinforce the need for non-equilibrium modeling approaches that reflect the physical reality of high-gradient, multiphase plume environments—particularly in the context of emerging propulsion technologies for which empirical data remain limited.

What this work does not aim to provide is a detailed treatment of turbulence effects, which are known to influence entrainment and dilution in the far field. These effects will be addressed in future extensions of the model, which will incorporate unsteady and turbulence-resolving simulations. Likewise, while the code base supports geometrically detailed nozzle designs, this study deliberately focuses on generic configurations to isolate phase-change mechanisms.

This work also does not claim to introduce fundamentally new physical models; rather, it combines and adapts well-established techniques from steam condensation and cloud microphysics into a unified numerical framework applicable to aircraft exhaust. By resolving the complex interplay between composition, thermodynamics, and nucleation physics in the near field, the model provides insight into the interface region between engine and atmosphere—a region that remains underexplored but is essential for setting the initial conditions of contrail formation.

Importantly, the modeling framework presented here is directly aligned with ongoing experimental work using a laboratory-scale contrail test rig. The same numerical approach will be applied to simulate these experiments, enabling direct validation of key phase-change mechanisms under controlled conditions. This combined computational–experimental platform is expected to provide a robust foundation for advancing contrail prediction capabilities and informing low-impact engine design.

By bridging detailed thermodynamic modeling and numerical simulation this work aims to contribute to the development of scalable, physically consistent tools for assessing and mitigating aviation’s non-CO$_2$ climate impacts. The results illustrate the limitations of equilibrium models and emphasize the importance of capturing nucleation pathways and interphase dynamics during early plume evolution. It is hoped that the presented framework will support future efforts toward more accurate assessments of contrail formation across a broad range of propulsion concepts.

\bibliography{references}

\end{document}